\documentclass[final,english]{bullsrsl}[2022/06/15]

\usepackage[latin1]{inputenc}
\usepackage[T1]{fontenc}
\usepackage{multirow}
\hypersetup{
    colorlinks=true,
    linkcolor=blue,
    filecolor=blue,      
    urlcolor=magenta,
    citecolor=blue,
}
\usepackage{natbib} 
\usepackage{graphicx}
\newcommand{\orb}{${\Omega}$\,} 
\newcommand{\twoorb}{${2\Omega}$\,} 
\newcommand{\s}{${\omega}$\,} 
\newcommand{\twosp}{${2\omega}$\,} 
\newcommand{\be}{${\omega-\Omega}$\,} 
\newcommand{\twobe}{${2(\omega-\Omega)}$\,} 
\newcommand{\po}{P$_{\Omega}$\,} 
\newcommand{\ps}{P$_{\omega}$\,} 
\newcommand{\ptwoo}{P$_{2\Omega}$\,} 
\newcommand{\ptwos}{P$_{2\omega}$\,} 
\newcommand{\pb}{P$_{\omega-\Omega}$\,} 
\newcommand{\ptwob}{P$_{2(\omega-\Omega)}$\,} 

\begin{document}
\title{A preliminary timing analysis of two intermediate polars: UU Col and Swift J0939.7-3224}

\author[affil={1,2}, corresponding]{Nikita} {Rawat}
\author[affil={1}]{J. C.} {Pandey }
\author[affil={2}]{Arti} {Joshi}
\author[affil={1}]{Srinivas M}  {Rao}
\author[affil={3}]{Micha\"{e}l} {De Becker}
\affiliation[1]{Aryabhatta Research Institute of observational sciencES (ARIES), Nainital 263001, India}
\affiliation[2]{Deen Dayal Upadhyaya Gorakhpur University, Gorakhpur 273009, India}
\affiliation[3]{Pontificia Universidad Cat\'{o}lica de Chile, Av. Vicu\~{n}a Mackenna 4860, 782-0436 Macul, Santiago, Chile}
\affiliation[4]{Space Sciences, Technologies and Astrophysics Research (STAR) Institute, University of Li\`{e}ge, Quartier Agora, 19c, All\'{e}e du 6 A\^{o}ut, B5c, B-4000 Sart Tilman, Belgium}
\correspondance{rawatnikita221@gmail.com}
\maketitle


%

\begin{abstract}
We present the preliminary timing analysis of confirmed intermediate polar UU Col and possible intermediate polar Swift J0939.7-3224 in the optical band with the help of long-term, high-cadence continuous photometry from \textit{Transiting Exoplanet Survey Satellite (TESS)}. For UU Col, we revise previously reported orbital and spin periods as 3.464 $\pm$ 0.005 h and 863.74 $\pm$ 0.08 s, respectively. Using the second harmonic of the beat frequency, the beat period is estimated as $\sim$928 s. These findings indicate that UU Col is a disc-fed dominated disc-overflow accretor. For J0939, we establish the spin period as 2671.8 $\pm$ 0.8 s and refine the provisionally suggested orbital period as 8.49 $\pm$ 0.03 h. The absence of beat frequency in J0939 signifies that it might be a pure disc-fed accretor; however, an X-ray study of this source will help to understand its true nature.
\end{abstract}


\section{Introduction}
The modest magnetic field strength (B $\sim$1-10 MG) subclass of magnetic cataclysmic variables (MCVs) is known as intermediate polars (IPs). These are semi-detached binaries where the magnetic field influences the accretion process that takes place in the system, as the matter is transferred from the companion (typically, a Roche-lobe filling late-type star) to the primary (white dwarf; WD). A full review of IPs is available in \cite{1983ASSL..101..155W, 1994PASP..106..209P,  1995ASPC...85..185H}. For IPs, the spin period of WD (\ps) is less than the orbital period (\po) of the binary system (generally, \ps/\po $\sim$0.1). A truncated accretion disc can form due to a moderately strong magnetic field in these systems. Therefore, accretion occurs via disc or stream or a combination of both. The mass accretion rate, the magnetic field strength of WD, and binary orbital separation are the key factors that determine which of the three accretion scenarios: disc-fed, stream-fed, and disc-overflow occurs in these systems. Because of the intricate interactions between the spin and orbital modulations, IPs are characterized by the existence of various periodicities in the X-ray and optical power spectra, which decide the governing accretion scenario. The sole presence of the spin frequency (\s) in the power spectrum depicts disc-fed accretion \citep{1995A&A...298..165K, 1996MNRAS.280..937N}, while for a pure stream-fed accretion, there is strong power at the lower orbital sideband of the spin frequency, i.e., beat (\be) frequency  \citep{1986MNRAS.218..695H, 1991MNRAS.251..693H, 1992MNRAS.255...83W, 1993MNRAS.265..316N}. Although \s frequency can also be present in the case of a stream-fed accretion if the magnetic poles are asymmetric in terms of polecap luminosities and size, the presence of 2\s-\orb ~frequency decides between a pure stream-fed and dic-fed accretion \citep{1992MNRAS.255...83W}. On the other hand, for a disc-overflow accretion, both frequencies (\s and \be) should be simultaneously present in the power spectrum \citep{1989ApJ...340.1064L, 1996ApJ...470.1024A, 1991MNRAS.251..693H, 1993MNRAS.265L..35H}. Disc-overflow accretion can further be described into three categories: disc-fed dominated disc-overflow accretor, stream-fed dominated disc-overflow accretor, and disc-overflow with equal dominance depending on whether the dominating frequency is \s,~\be ~or equal dominance of both, respectively. IPs are fascinating candidates for understanding the physics of magnetically controlled accretion. Therefore, we selected two sources, namely UU Col and Swift J0939.7-3224, from the IP catalogue of Koji Mukai \url{(https://asd.gsfc.nasa.gov/Koji.Mukai/iphome/catalog/alpha.html)}  with a motivation to ensure the IP nature and to inspect the nature of the accretion flow.

\subsection{UU Col}
UU Col (discovery name: RX J0512.2-3241) was identified as a soft X-ray intermediate polar by \cite{1996A&A...310L..25B}. From B and V band  CCD photometric data, they derived two periods of 3.45 $\pm$ 0.03 h and 863.5 $\pm$ 0.7 s, suggested to be \po and \ps, respectively. \cite{1996A&A...310L..25B} also found a very weak signal at 928.1 $\pm$ 0.9 s, which was proposed to be \pb. From X-ray data, \cite{2006A&A...454..287D} found the dominant frequency corresponding to the period of 863.3 $\pm$ 1.4 s (\ps)  and very weak variability at \pb $\sim$ 935 s. The orbital period in X-ray data was found to be 3.55 $\pm$ 0.56 h.
\cite{2010ApJ...724..165K} found that the polarization is modulated at the spin period of the WD in UU Col with amplitudes between $+$0.4\% and $-$0.4\% and $+$0.6\% and $-$0.7\% in the B band and I band, respectively. These values suggest that the polarized variability is most likely due to the cyclotron emission from two poles.

\subsection{Swift J0939.7-3224}
Swift J0939.7-3224 (hereafter J0939) was identified by \cite{2015AJ....150..170H} from \textit{Swift}-XRT and \textit{Swift}-UVOT images as a candidate MCV.  From radial-velocity periodogram analysis and time-series photometric analysis, \cite{2015AJ....150..170H} found two probable periods in their data sets: 8.51 $\pm$ 0.02 h and 2670 $\pm$ 7 s, which were provisionally suggested to be \po and \ps, respectively.  Apart from this, we are not aware of any other published results on this source.

\section{Observations and Data}
We have utilized \textit{TESS} data of both targets for the present study. The \textit{TESS} instrument consists of four wide-field CCD cameras, each with a field-of-view of 24\textdegree $\times$ 24\textdegree ~so that all cameras can image a region of the sky measuring 24\textdegree $\times$ 96\textdegree. \textit{TESS} observations are broken up into sectors, each lasting about 27.4 days and conduct its downlink of data while at perigee. The \textit{TESS} bandpass extends from 600 to 1000 nm with an effective wavelength of traditional Cousins I-band, i.e., 786.5 nm \cite[see][for details]{2015JATIS...1a4003R}. We downloaded the data from Mikulski Archive for Space Telescopes (MAST) data archive (\url{https://mast.stsci.edu/portal/Mashup/Clients/Mast/Portal.html}) with unique identification numbers `TIC 77841332' and `TIC 25611385' for UU Col and J0939, respectively. The \textit{TESS} observations of UU Col and J0939 were conducted at a cadence of 2 min from November 19, 2020, to  December 16, 2020 (sector 32) and February 9, 2021, to March 6, 2021 (sector 35), respectively.

\begin{figure}[h]
\centering
\includegraphics[width=\textwidth,height=12cm]{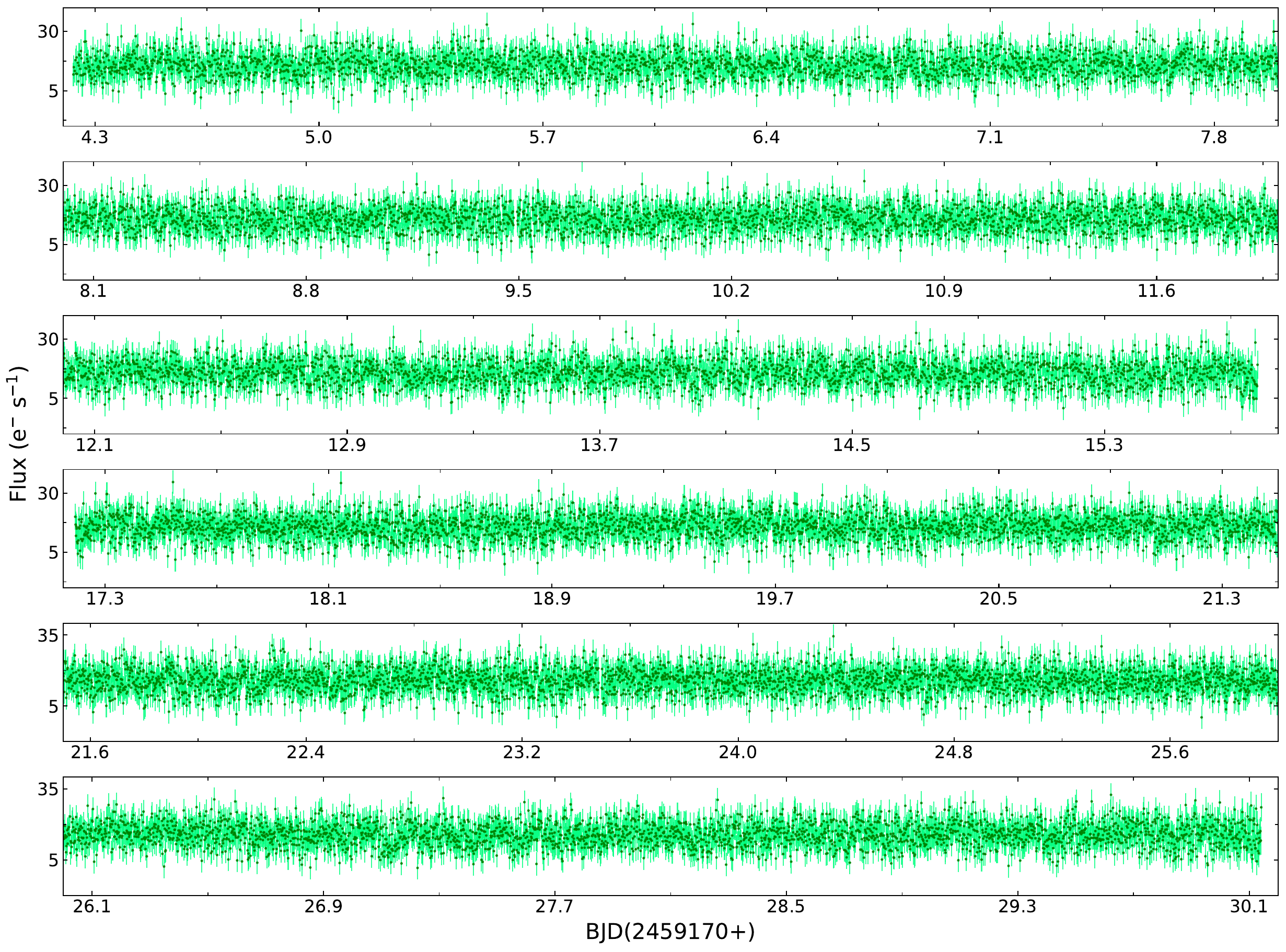}
\caption{\textit{TESS} light curve of UU Col spanning $\sim$26 d.}
\label{fig:1}
\end{figure}

\section{Timing Analysis}
\subsection{UU Col} 
\begin{figure}[h]
\centering
\includegraphics[width=\textwidth,height=8cm]{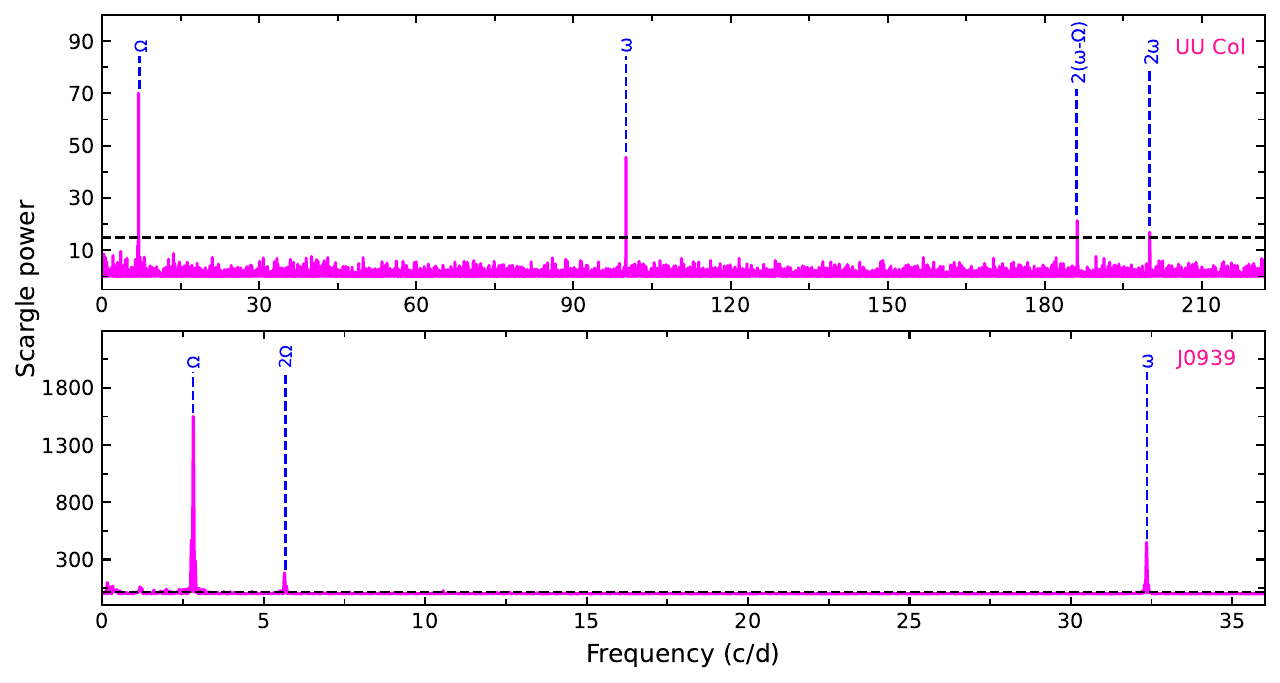}
\caption{Power spectra of UU Col and J0939, where all dominant frequencies are marked for clear visual inspection.}
\label{fig:2}
\end{figure}

Figure \ref{fig:1} depicts the \textit{TESS} light curve of UU Col, which, when closely examined, reveals short-term variations superimposed on long-term variability. In order to identify the periodicities in the data, we used Lomb-Scargle (LS) periodogram analysis \citep{1976Ap&SS..39..447L, 1982ApJ...263..835S} in which an error in the peak is obtained by computing the half-size of a single frequency bin centred on the peak and then converting it to period units. The power spectrum obtained from \textit{TESS} data is shown in the top panel of Figure \ref{fig:2}, where we have marked the position of all dominant frequencies. The significance of these frequencies was obtained by computing the false alarm probability \citep[FAP;][]{1986ApJ...302..757H} and the dashed line implies a 90 \% significance level, indicating that we are 90 \% confident that the present frequencies are intrinsic in origin and not due to noise. The power spectral analysis revealed the presence of \orb, \s, \twosp, and \twobe ~frequencies in the data and the corresponding periods are provided in Table \ref{tab:1}. The \orb ~frequency was found to be the dominant one, followed by \s and \twobe.

\begin{table}[h!]
\centering
\caption{Periods corresponding to dominant peaks in the LS power spectra of UU Col and J0939.}
\label{tab:1}
\bigskip

\begin{tabular}{cccccc} 
\hline
\textbf{Object} & \textbf{\po} (h) & \textbf{\ptwoo} (h) & \textbf{\ps} (s) & \textbf{\ptwos} (s) & \textbf{\ptwob} (s) \\
\hline
UU Col & 3.464 $\pm$ 0.005 & ..... & 863.74 $\pm$ 0.08 &  431.87 $\pm$ 0.02 & 464.00 $\pm$ 0.02 \\
J0939 & 8.49 $\pm$ 0.03 & 4.253 $\pm$ 0.008 & 2671.8 $\pm$ 0.8 & ..... & ..... \\

\hline
\end{tabular}
\end{table}

\subsection{J0939} 
\begin{figure}[h]
\centering
\includegraphics[width=\textwidth,height=12cm]{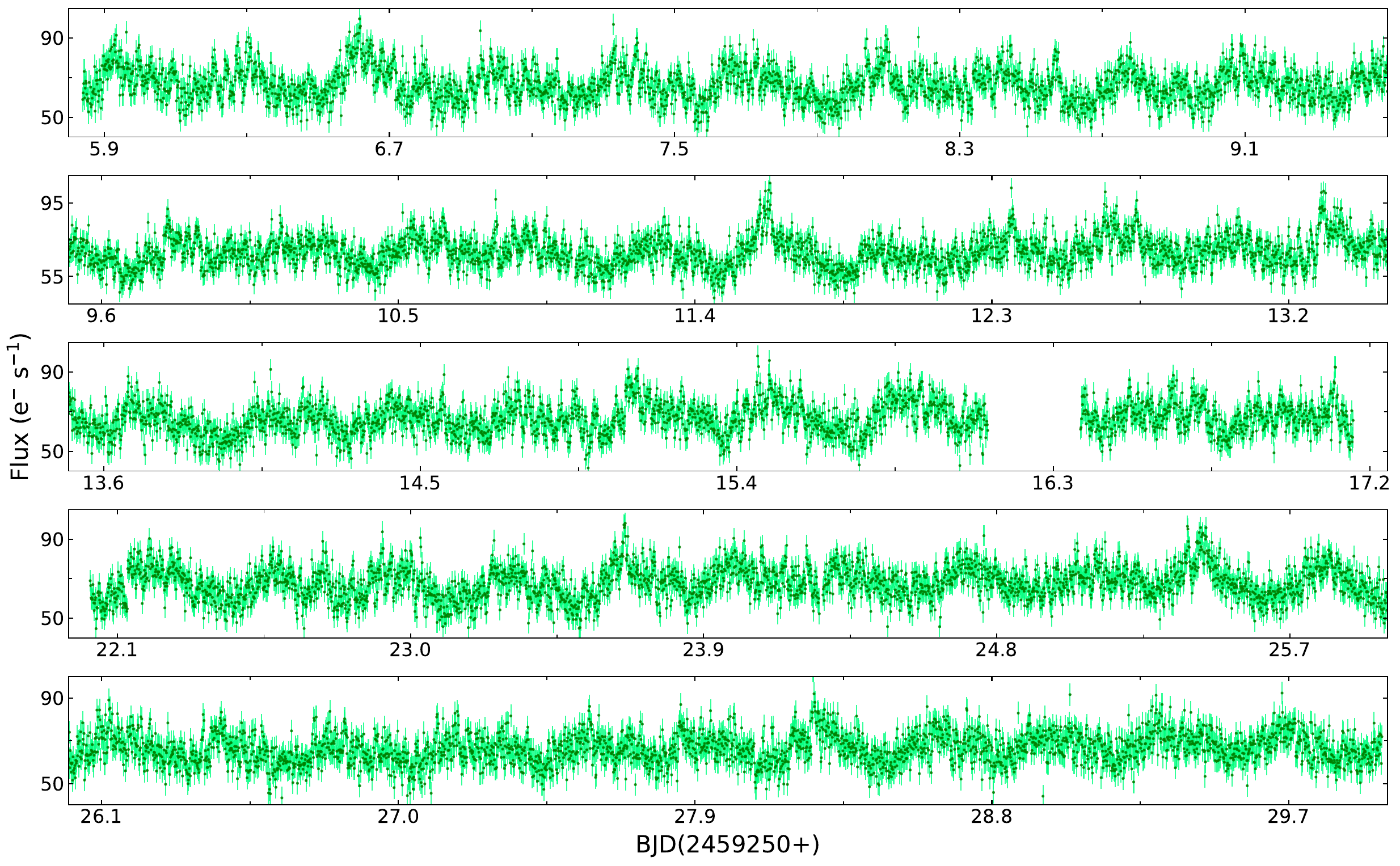}
\caption{\textit{TESS} light curve of J0939 spanning $\sim$24 d.}
\label{fig:3}
\end{figure}

In Figure \ref{fig:3}, we have shown the \textit{TESS} light curve of J0939 and a clear variability pattern is observed. As discussed earlier, we have performed the LS periodogram analysis and the bottom panel of Figure \ref{fig:2} shows the corresponding power spectrum. Similar to UU Col, FAP was calculated and the dashed horizontal line denotes a 90 \% significance level. The frequencies identified in the power spectrum are \orb, \twoorb, and \s. Due to the better time-cadence and longer observation duration in \textit{TESS}, we obtained more precise values of periods than available in the literature (see Table \ref{tab:1}). Using these values of \po and \ps, the inferred value of \pb ~period comes out to be $\sim$2927 s which corresponds to the frequency of $\sim$29.5 c/d. We do not see any peak near $\sim$29.5 c/d in our power spectrum, similar to the power spectrum of \cite{2015AJ....150..170H}. Further, the \orb frequency dominates the power spectrum over \s frequency.

\section{Discussion and Conclusions}
Using optical data from \textit{TESS}, we have performed a preliminary timing analysis of two IPs, UU Col and J0939. For UU Col, we refine previously determined orbital and spin periods, whereas, for J0939, we confirm that $\sim$2672 s is indeed the spin period of the WD. With \ps/\po $\sim$0.1, both targets fall in the category of intermediate rotators. The power spectrum plays a crucial role in determining the mode of accretion in these systems based on the presence of spin, beat, orbital, and other sideband frequencies and their amplitudes. UU Col was identified as a disc-overflow accreting IP by \cite{2006A&A...454..287D} since they found a very weak signal at \pb ~in their dataset. Contrary to \cite{1996A&A...310L..25B} and \cite{2006A&A...454..287D}, we have identified a clear peak at \twobe frequency in the power spectrum of UU Col, from which the inferred value of \pb ~comes out to be $\sim$928 s. This value is more precise than those obtained by \cite{1996A&A...310L..25B} and \cite{2006A&A...454..287D}. The presence of \twobe ~frequency in the power spectrum indicates that a part of the accreting material also flows through the stream along with the majority of accretion taking place via a disc. It also suggests that the field distribution might have an up-down symmetry. Therefore, we conclude that UU Col is a disc-fed dominated disc-overflow accretor (see Table \ref{tab:2} for different accretion scenarios based on frequencies present in the power spectrum). Furthermore, the presence of the spin period in the \textit{TESS} power spectrum of J0939 confirms that it is certainly an IP. The presence of \s ~frequency and absence of \be ~indicates that J0939 is a pure disc-fed accretor (see Table \ref{tab:2}), however, \cite{1999MNRAS.309..517F} have demonstrated the significance of stream extensions for the presence of \s frequency in the power spectrum. As the azimuthal extension of the accretion stream in the orbital plane increases, power shifts from \be ~to \s ~for stream-fed accretion.  Moreover, \cite{1992MNRAS.255...83W} have shown that stream-fed accretion can also produce a modulation at the spin frequency, in addition to that at the beat frequency, if there is an asymmetry between two magnetic poles. Hence, the presence of 2\s-\orb ~frequency in X-ray bands provides a clear identification between disc-fed and stream-fed systems and is only present in disc-less systems \citep{1992MNRAS.255...83W}. Therefore, X-ray observations of J0939 are needed to ascertain its nature further and see any change in governing accretion mechanism in the future. A variable disc-overflow accretion has been observed in many IPs, for example, TX Col \citep{1997MNRAS.289..362N, 2021ApJ...912...78R, 2021AJ....162...49L}, FO Aqr \citep{1996A&A...310L..25B, 2020ApJ...896..116L}, V2400 Oph \citep{2019AJ....158...11J}, V902 Mon and Swift J0746.3-1608 \citep{2022MNRAS.512.6054R}. In light of this, comprehensive time-resolved timing analysis of these targets is still required in the future.

\begin{table}[h!]
\centering
\caption{Accretion mechanisms and frequencies present in the optical power spectrum of UU Col and J0939.}
\label{tab:2}
\bigskip
\begin{tabular}{p{5cm} p{5cm} p{5cm}} 
\hline
\textbf{Accretion mechanism} & \textbf{Frequencies present} & \textbf{Remarks}  \\
\hline
Disc-fed accretion & \s ~and/or harmonics & Present in UU Col and J0939 \\
Stream-fed accretion & \be ~and/or harmonics & Present in UU Col \\
Disc-overflow accretion & \s, \be, ~and/or their harmonics & Present in UU Col\\
Disc-overflow with disc-fed dominance & \s, \be, ~and/or their harmonics and \s ~is the dominant frequency in comparison with \be & Present in UU Col\\
\hline
\end{tabular}
\end{table}


\begin{acknowledgments}
We thank the anonymous referee for reading our paper. NR and SMR acknowledge the Belgo-Indian Network for Astronomy and Astrophysics (BINA) consortium
for the local support to attend the 3$^{rd}$ BINA workshop.
 This work includes data collected with the \textit{TESS} mission, obtained from the MAST data archive at the Space Telescope Science Institute (STScI). Funding for the \textit{TESS} mission is provided by the NASA Explorer Program. STScI is operated by the Association of Universities for Research in Astronomy, Inc., under NASA contract NAS 5-26555.
\end{acknowledgments}

\begin{furtherinformation}

\begin{orcids}
\orcid{0000-0002-4633-6832} {Nikita} {Rawat}
\orcid{0000-0002-4331-1867} {J. C.} {Pandey}
\orcid{0000-0002-1303-6534} {Micha\"{e}l} {De Becker}
\end{orcids}

\begin{authorcontributions}
The corresponding author worked on the data and the core writing, while other authors contributed to the discussion, reviewing, and editing of the content.
\end{authorcontributions}

\begin{conflictsofinterest}
The authors declare no conflict of interest.
\end{conflictsofinterest}

\end{furtherinformation}

\bibliographystyle{bullsrsl-en}
\bibliography{ref}

\end{document}